\documentclass[twocolumn,epjc3]{svjour3} 
\usepackage[colorlinks,citecolor=blue,linkcolor=blue,anchorcolor=blue,filecolor=blue,urlcolor=blue]{hyperref}
\usepackage{graphicx}
\usepackage{ulem}
\usepackage{dcolumn}
\usepackage{bm}
\usepackage{amsmath}
\usepackage{amssymb}
\usepackage{amsfonts}
\usepackage{mathtools}
\usepackage{cite}
\usepackage{multicol}

\begin{document}

\title{Holographic hybrid stars with slow phase transitions}

\author{M.~Aleixo \thanksref{e1,addr1} \and C.~H.~Lenzi \thanksref{e2,addr1} \and M.~Dutra \thanksref{e3,addr1} \and O.~Louren\c{c}o \thanksref{e4,addr1} \and W. de Paula \thanksref{e5,addr1}}
\institute{
Departamento de F\'isica e Laborat\'orio de Computa\c c\~ao Cient\'ifica Avan\c cada e Modelamento (Lab-CCAM), Instituto Tecnol\'ogico de Aeron\'autica, DCTA, 12228-900, S\~ao Jos\'e dos Campos, SP, Brazil \label{addr1}}
\thankstext{e1}{e-mail: s.michaelaleixo@gmail.com} 
\thankstext{e2}{e-mail: chlenzi@ita.br}
\thankstext{e3}{e-mail: marianad@ita.br}
\thankstext{e4}{e-mail: odilon.ita@gmail.com}
\thankstext{e5}{e-mail: wayne@ita.br}



\date{\today}

\date{\today}
\maketitle
\begin{abstract}
The $D_3$-$D_7$ holographic model is used to describe the core of the hybrid star, composed by quark matter, while its crust is modeled from a hadronic relativistic mean field (RMF) model capable of reproducing low-energy nuclear physics data as well as some astrophysical observations. The $D_3$-$D_7$ brane configuration and the RMF model lead to an equation of state that is used to solve the Tolman-Oppenheimer-Volkoff equations. For different model parameters, the mass-radius diagram is presented. The conditions for the dynamic stability of stellar configurations are discussed, considering the radial oscillation criterion for hybrid stars with slow phase transitions. Strikingly, it is shown that the models generate stable star configurations with a core of quarks. We compare our results with NICER observational data for the pulsars PSR J0030+0451 and PSR J0740+6620 and show that the compact stars generated from this method fall within the corresponding observational regions.

\end{abstract}

\authorrunning{M. Aleixo et. al.}
\titlerunning{Holographic hybrid stars}

\maketitle

\section{Introduction}

Neutron stars (NS) are formed as remnants of supernovae depending on the mass of the progenitor star. The interior of a NS is not completely understood but it is expected that the pressure in its core can achieve very high values \cite{Lattimer:2015nhk}. Regarding its constitution, there are models that consider the presence of hyperons or $K^{-} $ condensate \cite{Glendenning:1991es,Bombaci:2008wg,Dexheimer:2008ax} and other predict that the pressure becomes so high that cause the hadronic dissociation into quarks \cite{Bodmer:1971we,Witten:1984rs,Terazawa:1978ni,Lattimer:2006xb}. If this occurs, the result is a hybrid star (HS), typically composed of a crust of hadronic matter and a core of deconfined quark matter. There is also a more extreme hypothesis for compact stars, suggesting that they could be entirely composed of quark matter, the quark stars \cite{Haensel:1986qb,Xu:2003xe,Lugones:2015bya,Lourenco:2021lpn,Chu:2023rty,Aleixo:2023lue}. Indeed, since the composite of the NS impacts observational properties, as the mass-radius~(MR) relation, the NS physics enables the study of dense matter under extreme conditions. It is also worth to mention that the gravitational waves \cite{LIGOScientific:2017ync,LIGOScientific:2017zic} detection from a binary NS merger by the LIGO-Virgo collaboration gives additional constraints in the NS physics.

The underlying theory to understand the interior of the NS is quantum chromodynamics (QCD). The difficulty is that the perturbative methods do not apply to NS physics, so the challenge is to deal with the QCD non-perturbative regime \cite{dePaula:2016oct,dePaula:2017ikc}. A promising approach is based on gauge/gravity duality, where, originally, the generating functional of the correlation functions of $\mathcal{N}= 4$ Super Yang-Mills (SYM) gauge theory is mapped to partition functions of type IIB superstring theory in  AdS$_{5} \times$ S$^{5}$ space \cite{Maldacena,WittenCorresp}. The duality main characteristic that makes it suitable for phenomenological application is that strongly coupled gauge theory is related to a weakly coupled classical gravity theory. Indeed, within the holographic perspective, there are new ways of describing linear confinement and spontaneous chiral symmetry breaking \cite{Klebanov:2000hb,Klebanov:2000nc,Maldacena:2000yy,Karch:2006pv,dePaula:2008fp,Bianchi:2010cy,dePaula:2009za,Ballon-Bayona:2023zal}. 

Holographic models also offer possibilities for the study of the QCD phase structure at finite temperature and density  \cite{DeWolfe:2010he,DeWolfe:2011ts,Knaute:2017opk,Ballon-Bayona:2020xls,dePaula:2020bte}. The application of holographic models is based on obtaining the corresponding equation of state (EoS) \cite{Hoyos:2016zke,AnnalaJokela,Jokela:2018ers,Mamani:2020pks,Hoyos:2021uff} and solving the Tolman-Oppenheimer-Volkoff equations (TOV)~\cite{Tolman,OppVolk}, which describe their hydrostatic equilibrium.

In this work, we aim to investigate the conditions that govern stellar stability of HS derived from equations of state computed using a Maxwell prescription. In this approach, we assume that the hadronic and quark phases are in direct contact, with only one of the two independent chemical potentials remaining continuous throughout the phase transition. The details will be provided in the following sections, but in summary, we use the D3-D7 top-down holographic model \cite{Karch:2002sh,Karch:2007br} for the quark phase, and a particular relativistic mean-field (RMF) model~\cite{PR464-113_2008,lattimer24,MDutra} parameterization, namely NL3*~\cite{NL3starorig}, for the hadronic sector. 

The transition can be classified as slow or rapid depending on the timescale of the conversion process from hadronic to quark matter relative to the timescale of perturbations caused by pressure fluctuations within the star. If the transformation timescale is significantly longer than that of the perturbations without mass transfer, the transition is considered slow; if it takes less time, it is deemed rapid \cite{LugonesRapidSlow,RapidSlowTrans} and some mass transfer can occur. 

The analysis of stellar stability depends on fast or slow transition being considered. In the first case, it is used the Bardeen-Thorne-Meltzer (BTM) criterion \cite{RapidBTM}, where stability is assessed by examining the concavity in the MR diagram. If the stellar mass increases as the central mass density increases, the star is considered stable. For slow transitions, the analysis is based on radial oscillations \cite{Vaeth1992,Chanmugam1977,Chandrasekhar1964}, which is a dynamical method also discussed in this work.

The D3-D7 holographic model was applied to the study of HS assuming fast phase transitions \cite{Hoyos:2016zke,AnnalaJokela,Hoyos:2021uff}. The authors obtained non-standard HS configurations, specifically the so-called second and third types, known as HS2 and HS3, respectively. The HS2 stars consist of a nuclear matter core surrounded by a crust of quark matter, also known as ``cross stars" \cite{CrossStars} and the HS3 stars feature a three-layer structure, with a nuclear matter core, a quark mantle and a nuclear crust \cite{AnnalaJokela}. Note that these configurations do not present a stable quark matter core. On the other hand, in the present work we show that it is possible to obtain a stable HS configuration with a quark matter core if one assume slow phase transitions, which impacts the stability analysis.

This work is organized as follows. In Section~\ref{D3D7}, the D3-D7 holographic model that stands for the quark phase is reviewed. Section~\ref{HadronicModels} presents the relativistic hadronic model that is used to build the HS crust. In Section~\ref{hsstability}, the criterion for dynamic stability through the analysis of radial oscillations is discussed. In Section~\ref{mrdiag}, the results obtained from the D3-D7/ NL3* model are provided. We present the MR diagram containing all stable segments and show that they fall within the observational regions \cite{PSRJ0740-1,PSRJ0740-2,PSRJ0030-1,PSRJ0030-2}. Additionally, we illustrate the stiffness of the hybrid equation of state through the gap in the sound speed squared at the phase transition for the range of constituent quark mass values studied. We also examine how the energy density varies within the isobaric regime and present a specific case of a pressure profile inside a $2.2$ solar mass star, which demonstrates a linear relationship at the phase transition points. In Section \ref{Conclusions}, our conclusions are presented. 

\section{The D3-D7 holographic model}
\label{D3D7}
The D3-D7 holographic model will be employed to describe the core of HS, where only quarks are present. This top-down approach is formulated by considering a stack of D3-branes intersecting with a probe D7-branes, which accounts for quark flavor. In the near-horizon limit of D3-branes, one obtains the overall spacetime, while the D7-brane, treated as a probe, captures the flavor dynamics of the quarks. The 10-dimensional space-time can be represented by the following table
\renewcommand{\arraystretch}{1.5} 
\setlength{\tabcolsep}{5pt}
\begin{table}[ht]
    \centering
    \vspace{0.2in}
    \label{tab:linelement2}
    \begin{tabular}{|c|c|c|c|c|c|c|c|c|c|c|}
        \hline
                & t         & $x_{1}$   & $x_{2}$    & $x_{3}$   & $x_{4}$   & $x_{5}$   & $x_{6}$   & $x_{7}$   & $x_{8}$  & $x_{9}$  \\ \hline
        D3   & $\bullet$ & $\bullet$ & $\bullet$   & $\bullet$  & -          & -          & -          & -         & -          & -         \\ \hline
        D7   & $\bullet$  & $\bullet$ & $\bullet$  & $\bullet$  & $\bullet$  & $\bullet$ & $\bullet$  & $\bullet$ & -         & -          \\ \hline
    \end{tabular}
\end{table}

\noindent and the 10-dimensional metric has the form
\begin{eqnarray}
    ds^{2} &=& \frac{u^{2}}{L^{2}} \; \eta_{\mu \nu} \; dx^{\mu} dx^{\nu} + \frac{L^{2}}{u^{2}} \; (d\bar{\rho}^{2} + \bar{\rho}^{2} \; d \Omega_{3}^{2}) \nonumber\\ 
    &+& \frac{L^{2}}{u^{2}} \; (dy^{2} + dz^{2}) \, ,
\end{eqnarray}
where $\eta_{\mu \nu}$ is the four-dimensional Minkowski metric, $L$ is the AdS radius and $u$ is the holographic coordinate, defined as $u^2 = \bar{\rho}^{2} + y^{2} + z^{2}$. The radial-spherical coordinate $\bar{\rho}$ and $\Omega_{3}$ belong to the $D_{7}$ brane worldvolume, given by $x_{4}$, $x_{5}$, $x_{6}$ and $x_{7}$. The coordinates $t$, $x_{1}$, $x_{2}$ and $x_{3}$ represent Minkowski space and form the set of dimensions shared by both D3 and D7-branes. Finally, the coordinates $x_{8}$ and $x_{9}$, represented by $y$ and $z$, correspond to the outer space that is not accessed by any brane.

The model operates in the `t Hooft limit, where the D7-brane wraps around an $AdS_{5} \times S^3$ \cite{Karch:2007br}, a subspace of the full $AdS_{5} \times S^5$ background. The low-energy dynamics are given by the Dirac-Born-Infeld (DBI) action
\begin{equation}
\label{actionD7}
    S_{D_{7}} = - N_{f} T_{D_{7}} \int  d^{8} \xi \; e^{- \phi} \; \sqrt{- \det{(g_{\mu \nu} + 2 \pi \alpha'} F_{\mu \nu})} \, ,
\end{equation}
\noindent where $N_{f}$ denotes the number of flavors and $T_{D7}$ is the tension of the D7-brane, which is expressed in terms of fundamental string parameters. The scalar field $\phi$ corresponds to the dilaton, while $g_{\mu\nu}$ is the induced metric on the worldvolume of the D7-brane. The parameter $\alpha'$ is the inverse of the string tension. The field strength $F$ is associated with a U(1) gauge field $A_{\mu}$, whose only non-vanishing component is the temporal one, $A_t(\bar{\rho})$.

The DBI action provides the following Lagrangian density
\begin{equation}
    \mathcal{L}_{DBI} = - \mathcal{N} \; \bar{\rho}^{3} \; \sqrt{1 + z^{' 2} - A^{'2}_{t}} \, ,
\end{equation}
\noindent where $\mathcal{N} = \frac{\pi^{2}}{2}N_{f} T_{D_{7}}$. For simplicity, since the term $2 \pi \alpha'$ is actually a constant, it can be absorbed into the gauge field.

The variations of the Lagrangian density with respect to $z$ and $A_{\mu}$ are zero. Consequently, two conserved quantities, $c$ and $d$, can be derived, given by
\begin{equation}
    c \equiv - \frac{1}{\mathcal{N}} \frac{\partial \mathcal{L}_{DBI}}{\partial z'} = \frac{\bar{\rho}^{3} \; z'}{\sqrt{1 + z^{' 2} - A^{'2}_{t}}} \, ,
\end{equation}
\begin{equation}
    d \equiv \frac{1}{\mathcal{N}} \frac{\partial \mathcal{L}_{DBI}}{\partial A'_{t}} = \frac{\bar{\rho}^{3} \; A'_{t}}{\sqrt{1 + z^{' 2} - A^{'2}_{t}}} \, .
\end{equation}
The two conserved quantities can be identified using the holographic dictionary, which states that under asymptotic conditions, $c$ and $d$ correspond to the constituent quark mass $m$ and the chemical potential $\mu$, respectively. It is important to note that this present quark holographic model adopts a flavor-independent approach. The constituent quark mass $m$ is the only free parameter in this model and will be used to construct different scenarios of HS.

The action (\ref{actionD7}) contains divergences. To address these, one can perform a regularization procedure at zero temperature by subtracting a specific term to obtain the regularized on-shell action \cite{Karch:2007br,Karch:2009zz}. This process allows for the final determination of the free energy density, which is composed of two distinct parts:
\begin{equation}
    \Omega = \Omega_{\mathcal{N} = 4} + \Omega_{\text{flavor}} \, . 
\end{equation}
The first part, $\Omega_{\mathcal{N} = 4}$, corresponds to the color contribution and can be neglected in the zero-temperature limit \cite{Mateos:2006}. The second part, $\Omega_{\text{flavor}}$, represents the flavor contributions, which can be expressed as \cite{Hoyos:2016zke}:
\begin{equation}
    \Omega_{\text{flavor}} = - \frac{3}{4 \pi^{2}} (\mu_{q}^{2} - m^{2})^{2} \, ,
\end{equation}
where the color and flavor numbers were set to three.

The free energy can be identified with the pressure $p_q$, thermodynamically expressed as $\Omega_{\text{flavor}} = -p_q$. By applying a Legendre transform, the energy density can be expressed as $\varepsilon = \mu_{q} \; \frac{\partial p}{\partial \mu_{q}} - p_q$, where $\mu_{q}$ is the chemical potential associated with the quark. The equation of state for the quark phase is \cite{AnnalaJokela}

\begin{equation}
    \varepsilon_q = 3p_q + \frac{2 \sqrt{3} \; m^{2}}{\pi} \; \sqrt{p_q} \, .
\end{equation}

\section{Relativistic mean-field model}
\label{HadronicModels}

In this work, the hadronic phase of the HS is described by a phenomenological finite-range relativistic mean-field model constructed by means of the Quantum Field Theory (hadrons as degrees of freedom). Specifically, such a model describes the nucleon-nucleon interaction in terms of the meson-exchange mechanism. A typical Lagrangian density used as the starting point for describing the hadronic system is given by~\cite{MDutra,lattimer24,PR464-113_2008}
\begin{align}
\mathcal{L}_{had} &= \bar{\psi}(i\gamma^\mu\partial_\mu - M_{\mbox{\tiny nuc}})\psi + g_\sigma\sigma\bar{\psi}\psi 
- g_\omega\bar{\psi}\gamma^\mu\omega_\mu\psi
\nonumber \\ 
&- \frac{g_\rho}{2}\bar{\psi}\gamma^\mu\vec{b}_\mu\vec{\tau}\psi
+\frac{1}{2}(\partial^\mu \sigma \partial_\mu \sigma - m^2_\sigma\sigma^2)
- \frac{A}{3}\sigma^3 
\nonumber\\
& - \frac{B}{4}\sigma^4  -\frac{1}{4}F^{\mu\nu}F_{\mu\nu} 
+ \frac{1}{2}m^2_\omega\omega_\mu\omega^\mu 
-\frac{1}{4}\vec{B}^{\mu\nu}\vec{B}_{\mu\nu} 
\nonumber\\
&+ \frac{1}{2}m^2_\rho\vec{b}_\mu\vec{b}^\mu.
\label{lagdens}
\end{align}
The nucleon and meson fields are represented by $\psi$, $\sigma$, $\omega^\mu$, and $\vec{b}_\mu$, with their corresponding masses denoted as $M{\mbox{\tiny nuc}}$, $m_\sigma$, $m_\omega$, and $m_\rho$. The field strength tensors associated with the vector mesons are defined as $F_{\mu\nu} = \partial_\mu \omega_\nu - \partial_\nu \omega_\mu$ and $\vec{B}_{\mu\nu} = \partial_\mu \vec{b}_\nu - \partial_\nu \vec{b}_\mu$. The model includes several free parameters, specifically the coupling constants $g_\sigma$, $g_\omega$, $g_\rho$, as well as the self-interaction coefficients $A$ and $B$. In this work, we adopt the parameter set from the NL3* model~\cite{NL3starorig}, selected due to its ability to accurately reproduce fundamental nuclear properties such as ground-state binding energies, charge radii, and giant monopole resonances. This parameterization has been validated against experimental data from various spherical nuclei, including $^{16}\rm O$, $^{34}\rm Si$, $^{40}\rm Ca$, $^{48}\rm Ca$, $^{52}\rm Ca$, $^{54}\rm Ca$, $^{48}\rm Ni$, $^{56}\rm Ni$, $^{78}\rm Ni$, $^{90}\rm Zr$, $^{100}\rm Sn$, $^{132}\rm Sn$, and $^{208}\rm Pb$. Additionally, it effectively describes the macroscopic properties of neutron stars. A comprehensive analysis involving over 400 other RMF model parameterizations is provided in Ref.~\cite{brett-jerome}.

By using the mean-field approximation~\cite{MDutra,PR464-113_2008}, we find the following field equations of the model
\begin{align}
m^2_\sigma\,\sigma &= g_\sigma \rho_s - A\sigma^2 - B\sigma^3, 
\\
m_\omega^2\,\omega_0 &= g_\omega \rho, 
\\
m_\rho^2\,b_{0(3)} &= \frac{g_\rho}{2}\rho_3, 
\\
[\gamma^\mu (&i\partial_\mu - g_\omega\omega_0 - g_\rho b_{0(3)}\tau_3/2) - M^*]\psi = 0,
\end{align}
with $\tau_3=1$ ($-1$) for protons (neutrons). The densities read 
\begin{align}
\rho_s &={\rho_s}_p+{\rho_s}_n,
\\
\rho &=\rho_p + \rho_n,
\\
\rho_3 &=\left<\bar{\psi}\gamma^0{\tau}_3\psi\right> = \rho_p - \rho_n =(2y_p-1)\rho, 
\\
\rho_{s_{p,n}} &= \left<\bar{\psi}_{p,n}\psi_{p,n}\right> =
\frac{M^*}{\pi^2}\int_0^{k_{F_{p,n}}} \hspace{-0.5cm}\frac{k^2dk}{(k^2+M^{*2})^{1/2}},
\end{align}
The vector densities are given in terms of the respective Fermi momenta, namely, ${k_F}_{p,n}=(3\pi^2\rho_{p,n})^{1/3}$, and the proton fraction is $y_p=\rho_p/\rho$. One of the effects of the attractive interaction induced by the $\sigma$ field is the change the nucleon mass by making it an in-medium quantity written~as
\begin{eqnarray}
M^* = M_{\mbox{\tiny nuc}} - g_\sigma\sigma.
\end{eqnarray}
In the above expressions, $\sigma$, $\omega_0$, $b_{0(3)}$ are the mean-field values of the respective mesonic fields. 

Concerning the thermodynamical quantities of the system, the energy density and pressure are obtained from the energy-momentum tensor, $T^{\mu\nu}$, calculated from Eq.~\eqref{lagdens}. These expressions read
\begin{align} 
\varepsilon_{had} &= \frac{1}{\pi^2} \int_0^{{k_F}_{p}}\hspace{-0.1cm}k^2({k^{2}+M^{*2}})^{1/2}dk 
+ g_{\omega} \omega_{0} \rho + \frac{g_{\rho}}{2} b_{0(3)}\rho_3
\nonumber\\
&+ \frac{1}{\pi^2} \int_0^{{k_F}_{n}}\hspace{-0.1cm}k^2({k^{2}+M^{*2}})^{1/2}dk
+ \frac{m_{\sigma}^{2} \sigma^{2}}{2} +\frac{A\sigma^{3}}{3}  
\nonumber\\
&+\frac{B\sigma^{4}}{4} -\frac{m_{\omega}^{2} \omega_{0}^{2}}{2} - \frac{m_{\rho}^{2} b_{0(3)}^{2}}{2},
\label{eden}
\end{align}
and
\begin{align}
&p_{had}= \frac{1}{3\pi^2} \int_0^{k_{F\,{p}}}\hspace{-0.5cm}  
\frac{k^4dk}{\left({k^{2}+M^{*2}}\right)^{1/2}}
+ \frac{m_{\omega}^{2} \omega_{0}^{2}}{2} + \frac{m_{\rho}^{2} b_{0(3)}^{2}}{2}
\nonumber\\
&=\frac{1}{3\pi^2} \int_0^{k_{F\,{n}}}\hspace{-0.5cm}  
\frac{k^4dk}{\left({k^{2}+M^{*2}}\right)^{1/2}}
-\frac{m_{\sigma}^{2} \sigma^{2}}{2} - \frac{A\sigma^{3}}{3} - \frac{B\sigma^{4}}{4}.
\label{press}
\end{align}

In order to fully describe the stellar matter on the hadronic side, we take into account the beta equilibrium condition in a system with electrons and muon added to the hadrons. They are given by 
\begin{align}
\rho_p - \rho_e  = \rho_\mu,
\end{align}
and 
\begin{align}
\mu_n - \mu_p = \mu_e=\mu_\mu.
\end{align}
In this context, the expressions for energy density and pressure are modified to
\begin{align}
\varepsilon_\beta &= \varepsilon_{had} + \frac{\mu_e^4}{4\pi^2}
+ \frac{1}{\pi^2}\int_0^{\sqrt{\mu_\mu^2-m^2_\mu}}\hspace{-0.6cm}dk\,k^2(k^2+m_\mu^2)^{1/2},
\end{align}
and
\begin{align}
p_\beta= p_{had} + \frac{\mu_e^4}{12\pi^2}
+\frac{1}{3\pi^2}\int_0^{\sqrt{\mu_\mu^2-m^2_\mu}}\hspace{-0.5cm}\frac{dk\,k^4}{(k^2+m_\mu^2)^{1/2}}.
\end{align}
The final terms in the above equations account for the thermodynamic contributions of massless electrons and muons, where the muon mass is $m_\mu=105.7$~MeV. The chemical potentials of these leptons, $\mu_e$ and $\mu_\mu$, are related to their respective number densities through the relations $\rho_e=\mu_e^3/(3\pi^2)$ and $\rho_\mu=[(\mu_\mu^2 - m_\mu^2)^{3/2}]/(3\pi^2)$.

Furthermore, the hadronic part of the compact star described in this work is divided into two regions: the outer crust and the inner crust (IC). The outer crust is described using the equations derived by Baym, Pethick, and Sutherland (BPS)~\cite{bps}, which are used within the density range of $6.3\times10^{-12}\,\mbox{fm}^{-3}\leqslant\rho\leqslant2.5\times10^{-4}\,\mbox{fm}^{-3}$. For the inner crust, we assume a polytropic equation of state relating pressure and energy density as 
\begin{align}
p_{\rm IC}=\mathcal{A}+\mathcal{B}\varepsilon_{\rm IC}^{4/3}
\end{align}
This expression is matched to both the BPS equation for the outer crust and the equation of state given by the RMF model. In particular, the connection between the inner crust and the matter described by the RMF model occurs at the core-crust transition, where the transition pressure and energy density are determined using the thermodynamical method\cite{tm1,tm2,tm4}.

\section{Results}

\subsection{Hadron-quark phase transition and stability analysis}
\label{hsstability}

We develop the hybrid EoS under the assumption that the interface between hadrons and quarks is a sharp discontinuity, modeled by means of the Maxwell construction. In other words, at zero temperature, the pressure and chemical potential remain equal on both sides of the phase transition boundary. With this goal, we combine various parameterizations of the holographic model D3-D7, described in Sect.~\ref{D3D7}, with a fixed RMF model shown in Sect. \ref{HadronicModels} (NL3*). In Table \ref{tab:exemplo_tabela}, we can observe the parameters related to phase transitions, including the jump in energy density $\Delta\varepsilon_t$, the transition pressure $p_t$, and the energy density at which the phase transition begins $\varepsilon_t$, for different values of the constituent quark mass $m$.

\begin{table}[!htb]
\centering
\caption{Transition data for slow-stable configurations of HS using the  D3-D7/NL3* model. The input is the constituent quark mass $m$. The following quantities are listed: $\varepsilon_{t}$ indicates the density energy of the hadronic part where the phase transition starts, $p_{t}$ is the corresponding transition pressure and $\Delta\varepsilon_{t}$ is the energy gap between the two phases. $\Delta \varepsilon_{t}$, $\varepsilon_{t}$ and $p_{t}$ are given in MeV/fm$^3$.}
    \vspace{0.2in}
    \label{tab:exemplo_tabela}
\begin{small}
    \begin{tabular}{|c|c|c|c|}
        \hline
        $m$ (MeV) & $\Delta \varepsilon_{t}$ & $\varepsilon_{t}$ & $p_{t}$ \\ \hline
        300   & 241.46   & 403.24 & 97.18    \\ \hline
        320   & 440.75   & 472.41 & 142.25   \\ \hline
        340   & 643.96   & 531.44 & 183.79   \\ \hline
        350   & 749.44   & 558.87 & 204.17   \\ \hline
        380   & 1091.40   & 640.59 & 265.36   \\ \hline
        390   & 1213.33   & 668.30 & 286.00   \\ \hline
    \end{tabular}
\end{small}
\end{table}

\begin{figure}[!htb]
 \centering
\includegraphics[trim=3cm 1cm 1cm 1cm, clip, scale=0.4]{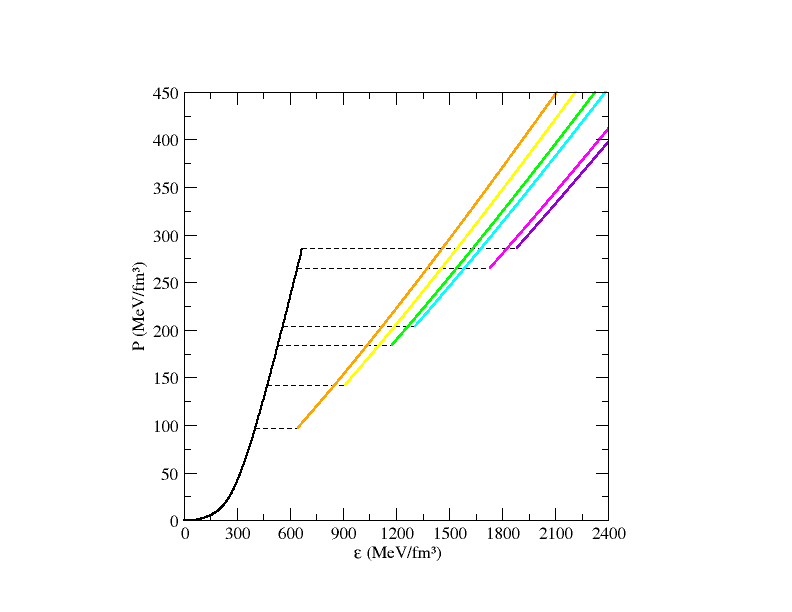} 
\caption{Pressure as a function of the energy density for the D3-D7/NL3* model (Maxwell construction implemented). The black curve is the NL3* hadronic model. The different values of the constituent quark mass in the curves representing the holographic model are the following: $m = 300$~MeV~(orange), $m = 320$~MeV~(yellow), $m = 340$~MeV~(green), $m = 350$~ MeV~(cyan), $m = 380$~MeV~(pink), and $m = 390$~MeV~(violet).}
\label{figPxENL3*}
\end{figure}

\begin{figure}[!htb]
 \centering
\includegraphics[trim=3cm 1cm 1cm 1cm, clip, scale=0.4]{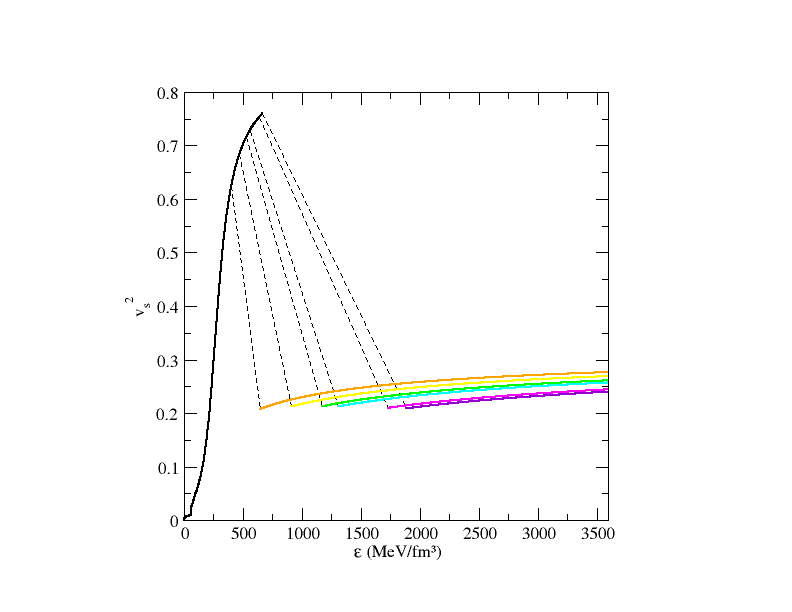} 
\caption{Squared sound speed as function of the energy density for the D3-D7/NL3* model. Same notation as in Fig.~\ref{figPxENL3*}.}
\label{figSqSSxENL3*}
\end{figure}

In Fig. \ref{figPxENL3*} we display the plot regarding the equations of state, with hadron-quark phase transition implemented, for different parameterizations of the holographic D3-D7 model. This graph illustrates the effect of increasing the constituent quark mass on the transition pressure and the energy density gap. In particular, both quantities increase as long as the constituent quark mass grows.
In Fig.~\ref{figSqSSxENL3*} we present the speed of sound $v_s^2=\partial p/\partial \varepsilon$ as a function of the energy density $\varepsilon$.
It is an indicator if causality is fulfilled, i.e. $v_s^2< 1$. For the range of parameters of Fig.~\ref{figSqSSxENL3*}, the present holographic hybrid model satisfies the causality criterion, explicitly showing that the speed of sound is higher in the hadronic phase.

The hydrostatic equilibrium for a spherically symmetric object is ensured by solving the TOV equations, which in natural units are given by
\begin{eqnarray}
\frac{dp (r)}{dr} &=& -\frac{ M (r) \, \varepsilon(r)}{r^{2}} \left(1 + \frac{4 \pi r^{3} p(r)}{M (r)} \right) \left(1 + \frac{p(r)}{\varepsilon(r)} \right) \nonumber\\
&& \times \, \left(1 - \frac{2 M (r)}{r} \right)^{-1} \, , \label{TOV1}\\
   \frac{d M (r)}{dr} &=& 4 \pi r^{2} \varepsilon(r) \, , \label{TOV2}
\end{eqnarray}
\noindent where $p(r)$ and $\varepsilon(r)$ are the pressure and energy density profiles of the star, respectively, in terms of the radial coordinate $r$, and $M(r)$ function is the Misner-Sharp mass. To solve the TOV equations it is necessary to consider the following boundary conditions at the center ($r = 0$) and at the surface of the star ($r = R$): $M(r = 0) = 0$ and $p(R) = 0$.

In this work, to analyze the stability of HS we solve the equations of radial oscillations. This analysis focuses on determining whether radial oscillations can maintain the stellar stability, as originally proposed by Chandrasekhar \cite{ChandrasekharRO}. The differential equations governing the radial oscillations are \cite{ChanmuganRO}
\begin{equation}
    \frac{d \xi}{d r} = - \frac{1}{r} \left(3 \xi + \frac{\Delta p}{\Gamma p} \right) - \frac{d p}{d r} \frac{\xi}{(p + \varepsilon)} \, ,
    \label{firstRO}
\end{equation}
\begin{equation}
\begin{aligned}
\frac{d \Delta p}{d r} &= \xi \bigg( \omega^2 e^{\lambda - \nu} (p + \varepsilon) r  
- 4 \frac{d p}{d r}  
+ \left(\frac{d p}{d r}\right)^2 \frac{r}{(p + \varepsilon)}  \\
&\quad - 8 \pi e^{\lambda} (p + \varepsilon) p \; r \bigg)  
+ \Delta p \bigg( \frac{d p}{d r} \frac{1}{(p + \varepsilon)}  \\
&\quad - 4 \pi (p + \varepsilon) r e^{\lambda} \bigg)
\label{bigeqRO}
\end{aligned}
\end{equation} 
 where $\Gamma$ is the relativistic adiabatic index. The functions $\nu$ and $\lambda$ depend on the radial coordinate $r$. The perturbation of the pressure and the relative radial displacement, $\Delta p$ and $\xi$,
respectively, are assumed to have a time dependence of
$e^{i\omega t}$, where the $\omega$'s are the eigenfrequencies corresponding to the normal modes of vibration. Generally, in stable stellar configurations, this frequency satisfies the condition $\omega > 0$, and for a single-phase stars the last stable configurations (where $\omega = 0$) corresponds to the point where $\frac{\partial M}{\partial \rho}$ changes the sign. However, in stellar models featuring a first-order phase transition with slow conversion at the interface, fundamental model can still have positive values beyond the maximum  mass point~\cite{cesar1,cesar2,cesar3}.

\subsection{Mass-radius diagrams}
\label{mrdiag}

Using the hybrid EoS obtained from the D3-D7/NL3* model, the hydrostatic equilibrium equations (\ref{TOV1})-(\ref{TOV2}) were solved for a range of constituent quark mass from $m = 330$ MeV to $m = 390$ MeV. The range of values chosen for the constituent quark mass are around $345$ MeV, which is the infrared mass function \cite{Castro:2023bij,Duarte:2022yur} obtained from lattice QCD calculations \cite{Oliveira:2018lln}. For a given central pressure $p_c$, the TOV equations were solved. 

\begin{figure}[!htb]
 \centering
\includegraphics[trim=3cm 1cm 1cm 1cm, clip, scale=0.4]{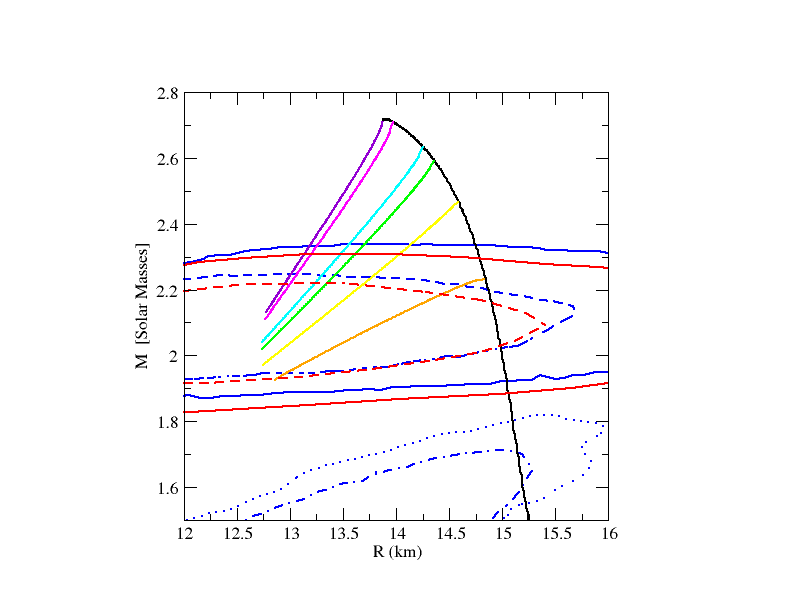} 
\caption{Mass-radius diagram for HS is presented using the D3-D7/NL3* model. The solid red and dashed curves represent the 95\% confidence intervals for masses and radii of the pulsar PSR J0740+6620 recently measured by NICER and XMM-Newton collaboration \cite{Dittmann:2024mbo}.
The blue curves are related to the 95\% confidence intervals for masses and radii measured by NICER, with the dashed and solid lines corresponding to the pulsar PSR J0740+6620 \cite{PSRJ0740-1,PSRJ0740-2}, while the dotted and dash-dotted lines refer to the pulsar PSR J0030+0451 \cite{PSRJ0030-1, PSRJ0030-2}. In all cases, the outer and inner lines corresponding to 1$\sigma$ and 2$\sigma$ intervals, respectively. Colors code: same notation as in Fig.~\ref{figPxENL3*}.}
\label{figMRNL3*}
\end{figure}

In Fig.~\ref{figMRNL3*}, it is presented the MR diagram for each constituent quark mass. All curves were computed until the last stable star taking into account slow phase transitions based on the positivity of the squared fundamental eigenvalue, $\omega_0^2$.
Remarkably, for all six parameterization, the holographic model predicts stable HS with mass higher than 2 solar masses, which is consistent with the analysis of the pulsars PSR J1614-2230, PSR J0348+0432 and PSR J0740+6620 \cite{Demorest:2010bx,Antoniadis:2013pzd,NANOGrav:2019jur}. Specifically, for $m=390$ MeV, the maximum HS mass reaches 2.7 $M_\odot$. Additionally, the model is compatible with recent NICER+XMM-Newton observational data \cite{Dittmann:2024mbo}. The  MR curves have stable segments that fall within the 95\% confidence intervals for the masses and radii of the pulsars PSR J0740+6620 \cite{PSRJ0740-1,PSRJ0740-2,NANOGrav:2019jur} and PSR J0030+0451 \cite{PSRJ0030-1,PSRJ0030-2}.

\begin{figure}[!htb]
 \centering
\includegraphics[trim=3cm 1cm 1cm 1cm, clip, scale=0.4]{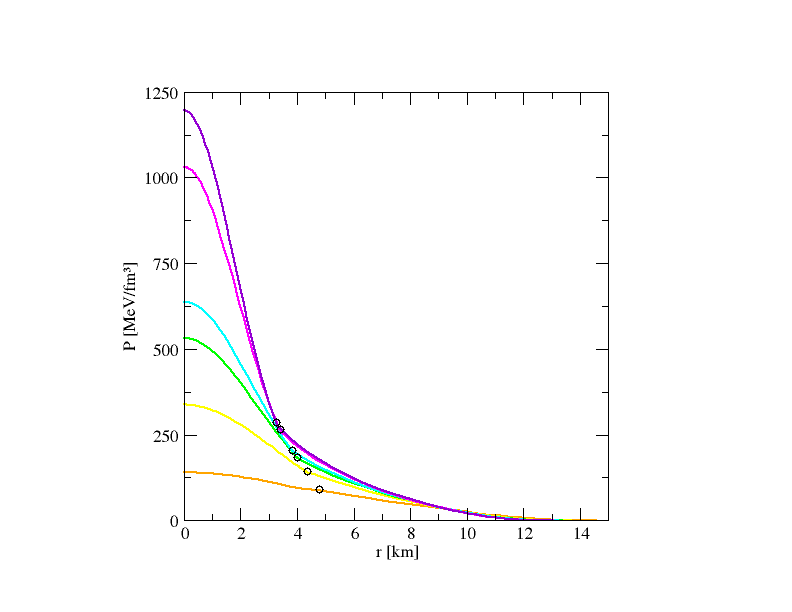} 
\caption{Hybrid star radial profiles for the 2.2 solar masses of each parameterization. Pressure versus radial coordinate. Colors code: same notation as in Fig.~\ref{figPxENL3*}. Black circles indicate where the phase transition occurs.}
\label{NL3*Profile}
\end{figure}

In Fig.~\ref{NL3*Profile} the radial profiles for 2.2 solar masses of each parameterization are depicted. 
The figure shows how pressure is distributed within the HS as a function of the radial coordinate. The maximum central pressure is obtained for $m = 390 $ MeV and the minimum is achieved for $m = 300$ MeV. It is observed a monotonic decreasing with the value of the constituent quark mass. In those curves are marked with a black circle to the radial position where the deconfinement hadron quark occurs for each parameterization. In Table~\ref{quarkcore}, the masses and radii of the quark core for each constituent quark mass are listed.
\begin{table}[!htb]
    \centering
    \caption{Quark core mass and radius for each parameterization for slow-stable configurations of $2.2 \; M_\odot$ HS using the D3-D7/NL3* model.}
    \vspace{0.2in}
    \label{quarkcore}
    \begin{tabular}{|c|c|c|}
        \hline
        $m$ (MeV) & $M_{core}$ ($M/M\odot$) & $r_{core}$ (km)  \\ \hline
        300   & 0.231   & 4.788   \\ \hline
        320   & 0.367   & 4.354   \\ \hline
        340   & 0.391   & 3.994   \\ \hline
        350   & 0.394   & 3.830   \\ \hline
        380   & 0.388   & 3.390   \\ \hline
        390   & 0.385   & 3.263   \\ \hline
    \end{tabular}
\end{table}

\section{Summary and concluding remarks}
\label{Conclusions}

In this work, we analyzed HS properties with a D3-D7/NL3* model. The quark phase was described by a D3-D7 brane configuration. In this holographic model, the constituent quark mass is a free parameter and, in the present study, we analyzed the HS properties in a range of $m=300$ MeV to $m=390$ MeV. The description of the hadronic phase was done by means of a particular parameterization of a relativistic mean-field model, namely, NL3*~\cite{NL3starorig}, that was shown to describe low-energy nuclear physics data while also aligning with certain astrophysical observations related to neutron stars~\cite{brett-jerome}. For the deconfinement hadron-quark we use the Maxwell construction in first order phase transitions.
We solved the TOV equations with appropriate boundary conditions, and the equations of radial oscillation to determine the stable regions on the MR diagram considering slow interface conversions.

The main outcome is that it is possible to have a stable HS with a core of quark matter consistent with the recent NICER +XMM-Newton observational data. Indeed, in Fig.~\ref{figMRNL3*}, the MR relation for a set of parameters was compared against pulsars observational data. We show that the stable hybrid star maximum mass increases with the constituent quark mass. In all six parameterizations, the HS maximum mass is higher than 2 solar masses, aligning with the analysis of the observational data of the pulsars PSR J0740+6620 \cite{PSRJ0740-1,PSRJ0740-2}, PSR J1614-2230 \cite{Demorest:2010bx} and PSR J0348+0432 \cite{Antoniadis:2013pzd}. Our calculations also illustrate that reaching 2.7 solar masses is achievable. The obtained $M(R)$ sequence of HS falls within the $95\%$ confidence level data analysis from the recent work of the NICER + XMM-Newton collaboration of the pulsar PSR J0740+6620 \cite{Dittmann:2024mbo} and from the NICER collaborations of the milisecond pulsars PSR J0740+6620 \cite{PSRJ0740-1,PSRJ0740-2,NANOGrav:2019jur} and PSR J0030+0451 \cite{PSRJ0030-1,PSRJ0030-2}.

In Table \ref{quarkcore} we can see the masses and radii of the quark core for hybrid stars of 2.2 solar masses. Raising the constituent quark mass leads to an increase in the quark core mass while reducing the radius. This means that the core becomes more compact. The same result can be seen when the pressure profile is analyzed, see Fig.~\ref{NL3*Profile}, where the hadron-quark interfaces can be identified in each case.

A desirable improvement to the holographic model would be incorporating the breaking of the SU(3) flavor symmetry in the quark phase. This would account for up, down, and strange quarks having distinct masses. Additionally, the model should include the presence of electrons to ensure beta equilibrium and electric charge neutrality in the quark sector.

\section*{Acknowledgements}
This work is a part of the project INCT-FNA proc. No. 464898/2014-5. M.A. acknowledges the partial support of the National Council for Scientific and Technological Development CNPq (Grant No. 400879/2019-0). C. H. L. is thankful to the São Paulo Research Foundation FAPESP (Grant No. 2020/05238-9) and to CNPq (Grants No. \\ 401565/2023-8 and 305327/2023-2). It is also supported by CNPq under Grants No. 307255/2023-9 (O.L.), No. 308528/2021-2 (M.D.). O.~L. and M.~D. also thank CNPq for the project No. 01565/2023-8~(Universal). W. d. P. acknowledges the partial support of CNPQ under Grant No. 313030/2021-9.

\bibliographystyle{epj}
\bibliography{references}

\end{document}